\documentclass[conference,10pt]{IEEEtran}
\IEEEoverridecommandlockouts
\def\BibTeX{{\rm B\kern-.05em{\sc i\kern-.025em b}\kern-.08em
    T\kern-.1667em\lower.7ex\hbox{E}\kern-.125emX}}

\usepackage[utf8]{inputenc}
\usepackage{cite}
\usepackage{amsmath,amssymb,amsfonts}
\usepackage{algorithmic}
\usepackage{graphicx}
\usepackage{textcomp}
\usepackage[caption=false]{subfig}
\usepackage[ruled,vlined]{algorithm2e}
\usepackage{epsfig}
\usepackage{float}
\usepackage{color}
\usepackage{balance}
\usepackage{bbm}
\usepackage{textcomp}
\usepackage{enumitem}
\usepackage{cite}
\usepackage{comment}

\newsavebox\myboxA
\newsavebox\myboxB
\newlength\mylenA

\def\mF{\mbox{$\mathbf{F}$}}
\def\mH{\mbox{$\mathbf{H}$}}

\def\mV{\mbox{$\mathbf{V}$}}

\def\mSigma{\mbox{$\mathbf{\Sigma} \kern .08em$}}
\def\mLambda{\mbox{$\mathbf{\Lambda} \kern .08em$}}

\def\b0{\text{\mbox{\boldmath $0$}}}

\def\bff{\text{\mbox{\boldmath $f$}}}

\def\bg{\text{\mbox{\boldmath $g$}}}
\def\bh{\text{\mbox{\boldmath $h$}}}
\def\bh{\text{\mbox{\boldmath $h$}}}

\def\bp{\text{\mbox{\boldmath $p$}}}

\def\bx{\text{\mbox{\boldmath $x$}}}

\def\e{\text{\varepsilon}}

\def\bv{\text{\mbox{\boldmath $v$}}}

\newcommand{\qedsymbol}{\hspace{\fill}\rule{1.5ex}{1.5ex}}

\IEEEoverridecommandlockouts 

\begin{document}

\title{Distributed Sum-Rate Maximization of Cellular Communications with Multiple Reconfigurable Intelligent Surfaces}

\author{Konstantinos D. Katsanos$^1$, Paolo Di Lorenzo$^{2,3}$, and George C. Alexandropoulos$^1$\\
$^1$Department of Informatics and Telecommunications, National and Kapodistrian University of Athens, Greece
\\$^2$Department of Information Engineering, Electronics, and Telecommunications, Sapienza University, Italy
\\$^3$ National Inter-University Consortium for Telecommunications (CNIT), Italy
\\e-mails: \{kkatsan, alexandg\}@di.uoa.gr, paolo.dilorenzo@uniroma1.it 
\thanks{This work was supported by the EU H2020 RISE-6G project under grant number 101017011.}  \vspace{-0.58cm}
}

\maketitle
\thispagestyle{empty}

\begin{abstract}
The technology of Reconfigurable Intelligent Surfaces (RISs) has lately attracted considerable interest from both academia and industry as a low-cost solution for coverage extension and signal propagation control. In this paper, we study the downlink of a multi-cell wideband communication system comprising single-antenna Base Stations (BSs) and their associated single-antenna users, as well as multiple passive RISs. We assume that each BS controls a separate RIS and performs Orthogonal Frequency Division Multiplexing (OFDM) transmissions. Differently from various previous works where the RIS unit elements are considered as frequency-flat phase shifters, we model them as Lorentzian resonators and present a joint design of the BSs' power allocation, as well as the phase profiles of the multiple RISs, targeting the sum-rate maximization of the multi-cell system. We formulate a challenging distributed nonconvex optimization problem, which is solved via successive concave approximation. The distributed implementation of the proposed design is discussed, and the presented simulation results showcase the interplay of the various system parameters on the sum rate, verifying the performance boosting role of RISs.
\end{abstract}

 \vspace{-0.30cm}
\section{Introduction} \label{sec:intro}
Reconfigurable Intelligent Surfaces (RISs) \cite{huang2019reconfigurable,di2019smart} are lately considered as a key enabling technology for future generation wireless communication networks, constituting a promising solution for realizing smart radio propagation environments \cite{RISE6G_COMMAG}. Many recent studies \cite{RIS_Overview} showcase that RISs are able to offer significant improvements in several performance requirements, such as energy efficiency as well as extended network coverage and connectivity for non-line-of-sight environments.

The concept of RIS-enabled smart wireless environments necessitates the efficient orchestration of multiple RISs \cite{Samarakoon_2020,Huang2021MultiHop,Stylianopoulos2022DeepCB}. By dynamically controlling the on-off states of each RIS, instead of considering that all of them are active, \cite{yang2020_resource} considered multiple RISs controlled by a single Base Station (BS) and focused on the resource allocation problem. In \cite{Marco21_multiRISMISO}, the authors presented a cooperative multi-RIS-assisted transmission scheme for millimeter-wave multi-antenna Orthogonal Frequency Division Multiplexing (OFDM) systems. Recently, an alternative model of delay-adjustable metasurfaces was proposed in \cite{an2021} for OFDM communications, where the RISs were controlled by a single BS, and the design objective was the transmit power allocation yielding the maximum achievable sum-rate performance. However, the vast majority of the up-to-date optimization frameworks considers centralized approaches, which require large overhead (in terms of time and power consumption) of control information exchange usually in a central network orchestrator.

In this paper, we investigate the impact of multiple distributed passive RISs, each one of them being controlled by a single-antenna BS that performs OFDM transmissions to its assigned single-antenna receiver, in the presence of interfering neighboring channels. Differently from other studies, we consider that each RIS meta-atom element is dictated by a frequency selective profile, according to the Lorentzian frequency response, which is more accurate for the considered OFDM-based multi-carrier modulations, as recently presented in \cite{katsanos2022wideband}. Based on the described system model, we formulate a design optimization problem focusing on the overall sum rate's maximization, and having as free parameters the power allocation over the Sub-Carriers (SCs) and the surfaces' reflection profiles. To solve the resulting problem, we develop a distributed optimization framework and propose a Successive Concave Approximation (SCA) algorithm which tackles the decoupling of the power allocation and the parameters of the Lorentzian responses. Through representative numerical results, the considerable gains of employing RISs for each BS-User Equipment (UE) pair are demonstrated, showcasing the notably improved sum rates offered compared to the case where RISs are not deployed.
\vspace{-0.05cm}

\textit{Notations:} Boldface lower-case and upper-case letters represent vectors and matrices, respectively. The transpose, Hermitian transpose, conjugate, and the real part of a complex quantity are represented by $(\cdot)^T$,  $(\cdot)^H$, $(\cdot)^*$, and $\Re\{ \cdot \}$, respectively, while $\mathbb{C}$ is the set of complex numbers, and $\jmath\triangleq\sqrt{-1}$ is the imaginary unit. The symbols $<\cdot,\cdot>$, $\mathbb{E}\{\cdot\}$, and $\circ$ denote the inner product, the statistical expectation, and the Hadamard product, respectively, while $\operatorname{diag}\{\mathbf{x}\}$ is defined as the matrix whose diagonal elements are the entries of $\mathbf{x}$. Finally, $[x]_+ \triangleq \max(0,x)$.

\section{System Model and Design Problem} \label{sec:sys_model} \vspace{-0.20cm}
We consider the design of a multi-user RIS-empowered and OFDM-based wireless system comprising of $Q$ communicating pairs. We focus on the downlink direction and assume that each BS sends information to its associated UE using a common set of physical resources, e.g., time and bandwidth. For the purpose of exposition, the BS and the UE are both equipped with a single antenna; the generalization to the multi-antenna case will be treated in the journal version of this work. Also, we assume that each BS can control an RIS, which is placed closely to it in order to enhance the communication with its UE.
The RISs are assumed to comprise $M$ passive reflecting elements, and are connected to a controller, which adjusts their pattern for desired signal reflection. We will henceforth refer to each BS-RIS-UE triplet as a “user”.
 \vspace{-0.20cm}
\subsection{Received Signal Model}
Similar to conventional OFDM-based communications, the total bandwidth is equally split into $K$ orthogonal SCs. Let $\bp_q$ = $[p_{q1},\ldots, p_{qK}]^T \in\mathbb{R}^K$, where each $p_{qk}\geq 0$ denotes the power allocated to the $k$-th SC by the $q$-th BS to transmit the unit-power signal $x_q[k]$ (i.e., $\mathbb{E}\{|x_q[k]|^2\} = 1$). We also assume that the total transmission power available at each BS $q$ is $P_q$. Thus, the power allocation must satisfy $\sum_{k=1}^K p_{qk}\leq P_q$ for each $q$-th BS. We consider a quasi-static block fading channel model for all channels involved, and focus on one particular fading block where the channels remain approximately constant. 

Let $y_q^d[k]$ denote the frequency-domain received signal at the $k$-th SC by the $q$-th UE through the direct channel $h_{qq}^d[k] \in \mathbb{C}$ between itself and its serving BS $q$. Moreover, there exists a multipath channel for the $q$-th BS-RIS-UE user, through which the signals transmitted by the $q$-th BS are reflected by its owned RIS before arriving at the $q$-th UE. Specifically, let $\bh_{qq}[k] \in \mathbb{C}^{M}$ denote the $q$-th BS-RIS channel at the $k$-th SC. Similarly, let $\bg_{qq}[k] \in \mathbb{C}^{M}$ denote the channel of the $q$-th RIS-UE link. At the $q$-th RIS, each reflecting element re-scatters the received signals with an independent reflection coefficient and we assume that the response of each $m$-th element (with $m=1,2,\ldots,M$) is modeled as a polarizable dipole whose frequency response at the $k$-th frequency bin, denoted by $\omega_k$, behaves according to the following Lorentzian form \cite{Shlezinger_2021_DMAs}, \cite{DSmith-2017PRA}:
\begin{equation}\label{eq:Lorentzian}
    \phi_q^m(\omega_k) = \frac{F_q^m \omega_k^2}{(\omega_q^m)^2 - \omega_k^2 + \jmath \kappa_q^m \omega_k},
\end{equation}
where $F_q^m,\, \omega_q^m$, and $\kappa_q^m$ are the element-dependent oscillator strength, angular frequency, and damping factor, respectively, which are the design parameters of each $q$-th RIS. Then, by letting $\boldsymbol{\phi}_{qk} \triangleq [\phi^1_{q}(\omega_k),\ldots,\phi^M_{q}(\omega_k)]^T \in \mathbb{C}^M$ denote the $q$-th RIS reflection coefficients vector and using the notation $\boldsymbol{\Phi}_q[k] \triangleq \operatorname{diag}\{\boldsymbol{\phi}_{qk}\}$, the frequency-domain  noiseless received signal part at each $q$-th UE from each serving $q$-th BS and $q$-th RIS can be expressed by the concatenation of the BS-RIS, RIS reflection, and RIS-UE channels as follows:
\begin{equation}\label{eq:RIS_signal}
    y[k] = \sqrt{p_{qk}}\big(h_{qq}^d[k] + \bg_{qq}^{T}[k] \boldsymbol{\Phi}_q[k] \bh_{qq}[k]\big)x_q[k],
\end{equation}

In the considered distributed scenario, no multiplexing strategy is imposed a priori so that, in principle, each user interferes with each other. In this case, the cross-channels among users are composed by the sum of a direct channel between the $j$-th BS and the $q$-th UE, which is denoted by $h_{jq}^d[k]$, and a reflected component due to the power reflected by the $j$-th RIS towards the $q$-th UE. Using similar arguments as in expression \eqref{eq:RIS_signal}, the overall received signal at each $k$-th SC at user $q$ is given by:  \vspace{-0.30cm}
\begin{equation} \label{eq:RIS_cross_signal}
    \begin{aligned}
    y_q[k] =& y[k] + \sum\limits_{j=1,j\neq q}^Q \sqrt{p_{jk}}\big(h_{jq}^d[k] \\
   &+ \bg_{jq}^T[k] \boldsymbol{\Phi}_j[k] \bh_{jj}[k]\big)x_j[k] + n_q[k],
\end{aligned}
\end{equation}
where $n_q[k]$ denotes the zero-mean additive white Gaussian noise with variance $\sigma_q^2$.

 \vspace{-0.20cm}
\subsection{Achievable Rate Performance} 
We focus on transmission schemes where no interference cancellation is performed and Multi-User Interference (MUI) is treated as additive colored noise from each UE receiver. To formulate the objective function, we assume that the UEs utilize wideband modulations. Moreover, based on the signal model in \eqref{eq:RIS_cross_signal} and letting $\bp \triangleq [\bp_1^T,\dots,\bp_Q^T]^T$ and $\tilde{\boldsymbol{\phi}} \triangleq [\boldsymbol{\phi}_1^T,\dots,\boldsymbol{\phi}_Q^T]^T$ with $\boldsymbol{\phi}_q \triangleq [\boldsymbol{\phi}_{q1}^T,\dots,\boldsymbol{\phi}_{qK}^T]^T$, the achievable rate in bits per second per Hertz (bps/Hz) for each $q$-th BS-RIS-UE link as a function of $\bp$, $\tilde{\boldsymbol{\phi}}$ can be expressed as:  \vspace{-0.10cm}
\begin{equation}\label{eq:rate_final}
    \mathcal{R}_q(\bp, \tilde{\boldsymbol{\phi}})=\sum_{k=1}^K\log_2\left(1\!+\!\frac{\lvert H_{qq}(\boldsymbol{\phi}_{qk}) \rvert^2 p_{qk}}{\sigma_q^2+ \sum\limits_{j\neq q} \lvert H_{jq}(\boldsymbol{\phi}_{jk}) \rvert^2 p_{jk}  }\right),
\end{equation}
where we have used the definitions $\forall$$q,j=1,2,\ldots,Q$:  \vspace{-0.10cm}
\begin{align}
    &H_{qq}(\boldsymbol{\phi}_{qk}) \triangleq h_{qq}^d + (\bg_{qq}^T \odot \bh_{qq}^T) \boldsymbol{\phi}_{qk}, \label{eq:equiv_channel_qq} \\
    &H_{jq}(\boldsymbol{\phi}_{jk}) \triangleq h_{jq}^d + (\bg_{jq}^T \odot \bh_{jj}^T) \boldsymbol{\phi}_{jk}, \label{eq:equiv_channel_jq}
\end{align}
for which we implicitly considered that each channel component is indexed by the frequency bin $\omega_k$ (e.g., $h_{qq}^d = h_{qq}^d[k]$). It is noted that the prelog factor $\frac{K}{K + K_{\rm CP}}$, with $K_{\rm CP}$ being the cyclic prefix length \cite{Zhang_2020_MIMO}, was neglected in \eqref{eq:rate_final} since it does not affect the optimization formulation and solution that will be presented in the sequel.
 \vspace{-0.10cm}
\subsection{Problem Formulation} \label{prob_form} 
Our goal in this paper is to distributively maximize a network utility function given by the sum rate of the users under power and RIS reflectivity constraints (in order to guarantee that the Lorentzian parameters of the RIS unit elements are such that the condition for passive beamforming is satisfied):  \vspace{-0.30cm}
\begin{align}\label{Problem:Max_rate}
\mathcal{OP}_1:\quad&\max_{\bp,\tilde{\boldsymbol{\phi}}}\;\;\sum_{q=1}^Q \, \mathcal{R}_q(\bp, \tilde{\boldsymbol{\phi}})\nonumber   \\  \vspace{-0.10cm}
& \quad \hbox{s.t.} \,\,\,\, \phi_{qk}^m = \frac{F_q^m \omega_k^2}{(\omega_q^m)^2 - \omega_k^2 + \jmath \kappa_q^m \omega_k},\nonumber\\
& \qquad \,\,\,\, p_{qk}\geq 0,\,\,\sum_{k=1}^K\;p_{qk} \leq P_q,\,\,|\phi_{qk}^m|\leq 1 \quad\forall k,m,q.\nonumber
\end{align} 
The latter objective function is not jointly concave in the power allocation and RIS parameters and, as a consequence, the problem has generally multiple local optima. In addition, the solution of $\mathcal{OP}_1$ requires, in general, a centralized approach. Nevertheless, we will show next that the solution can also be reached in a distributed fashion, by allowing a very limited exchange of control data among the involved UEs.
\section{Proposed Distributed Design}\label{sec:solution} 
Let $\bx_q=(\bp_q,\boldsymbol{\phi}_q)$ be the set of variables associated with user $q$. We also define $\bx_{-q}=(\{\bp_j\}_{j\neq q},\{\boldsymbol{\phi}_j\}_{j\neq q})$ as the set of all users' variables except the $q$-th one, and the set below: 
\begin{align}
    &\mathcal{X}_q=\Big\{ \bx_q=(\bp_q,\boldsymbol{\phi}_q)\;|\; p_{qk} \geq 0, \quad\forall\,  k,q,\nonumber\\
    &\qquad \qquad \sum_{k=1}^K\;p_{qk} \leq P_q,\quad |\phi_{qk}^m|\leq 1, \quad \forall k,m,q \Big\} \vspace{-0.10cm}
\end{align} 
which represents the feasible set for user $q$ in $\mathcal{OP}_1$. Finally, we define $\bx=\{\bx_q\}_{q=1}^Q$. Then, problem $\mathcal{OP}_1$ can be recast in the following compact form as: \vspace{-0.10cm}
\begin{align}\label{Problem:Max_rate2} 
\mathcal{OP}_2:\quad&\max_{\{\bx_q\}_{q=1}^Q }\;\;\sum_{q=1}^Q \, \mathcal{R}_q(\bx_q,\bx_{-q})\nonumber   \\
& \quad \hbox{s.t.} \qquad\bx_q \in\mathcal{X}_q, \qquad  q=1,2,\ldots,Q. \nonumber  \vspace{-0.20cm}
\end{align} 
We now proceed hinging on the methods from \cite{scutari2013decomposition,facchinei2015parallel}. In particular, for each user $q$, we build a (strongly) concave surrogate for the objective function in $\mathcal{OP}_2$ that can be computed thanks to limited exchange of information among users, and can be easily optimized in an iterative fashion. To this aim, we rewrite the sum-rate objective of this design problem in the following form: \vspace{-0.10cm}
\begin{equation}\label{eq:sum_rate}
    \overline{\mathcal{R}}(\bx_q,\bx_{-q})\,=\,\mathcal{R}_q(\bx_q,\bx_{-q})+\sum_{j\neq q} \mathcal{R}_j(\bx_q,\bx_{-q}). \vspace{-0.20cm}
\end{equation}  
Function \eqref{eq:sum_rate} is non-concave in both terms, due to the presence of MUI and the coupling between power allocation and RIS parameters. However, its structure leads naturally to a concavization having the following
form: i) at every iteration $t$, the nonconvex term $\mathcal{R}_q(\bx_q,\bx_{-q})$ is replaced by a strongly concave surrogate, say $\widetilde{\mathcal{R}}_q(\bx_q;\bx^t)$, which depends on the current iterate $\bx^t$; and ii) the term $\sum_{j\neq q} \mathcal{R}_j(\bx_q,\bx_{-q})$ is linearized around $\bx_q^t$. More formally, the proposed updating scheme reads: at every iteration $t$, each user $q$ solves the following strongly concave optimization problem:
\begin{equation*}\label{Surrogate_problem}
\mathcal{OP}_3:\quad\widehat{\bx}^t_q\,=\, \arg\max_{\bx_q \in\mathcal{X}_q} \, \widetilde{\mathcal{R}}_q(\bx_q;\bx^t)+<\boldsymbol{\pi}^t_q , \bx_q -\bx_q^t>, 
\end{equation*} \vspace{-0.25cm}
where we have defined the following functions:
\begin{align}
& \boldsymbol{\pi}^t_q = \sum_{j\neq q} \nabla_{\bx_q^*} \mathcal{R}_j(\bx_q,\bx_{-q})\Big|_{\bx_q=\bx_q^t} =  \sum_{j\neq q} \boldsymbol{\pi}^t_{qj} \label{eq:Prices}\\
& \widetilde{\mathcal{R}}_q(\bx_q;\bx^t)= \sum_{k=1}^K \log_2\left(1+\frac{|H^k_{qq}(\boldsymbol{\phi}^t_{qk})|^2 p_{qk}}{\sigma_{q}^2+ \sum\limits_{j\neq q} |H^k_{jq}(\boldsymbol{\phi}^t_{jk})|^2 p^t_{jk}  }\right)  \nonumber\\\vspace{.3cm}
&\quad+<\boldsymbol{\gamma}^t_q, \boldsymbol{\phi}_q-\boldsymbol{\phi}^t_q>-\frac{\tau}{2}\|\bp_q-\bp_q^t\|^2-\frac{\tau}{2}\|\boldsymbol{\phi}_q-\boldsymbol{\phi}_q^t\|^2 \label{eq:Surrogate}
\end{align}
\begin{equation}
    \boldsymbol{\gamma}^t_q= \nabla_{\bx_q^*} \mathcal{R}_q(\bx_q,\bx^t_{-q}) \Big|_{\bx_q=\bx^t_q}, 
\end{equation}
for all $q=1,\dots,Q$, and with $\tau>0$. Thanks to the concavization in \eqref{eq:Prices} and \eqref{eq:Surrogate}, the objective of $\mathcal{OP}_3$ is continuously differentiable, strongly concave, and preserves the first optimality conditions of (\ref{eq:sum_rate}) around the current iterate $\bx^t_q$. Under such conditions, any fixed point of the mapping $\{\widehat{\bx}^t_q\}_{q=1}^Q$ in $\mathcal{OP}_3$ is a local maximum of the sum-rate problem in $\mathcal{OP}_2$. The terms in (\ref{eq:Prices}) are often called interference prices in the literature \cite{scutari2013decomposition}, since their role is to quantify the amount of interference produced by the resource allocation (in our case, powers and RIS parameters) of user $q$ towards other users $j\neq q$. Taking into account interference prices into the overall optimization help maximizing the social sum-rate utility function thanks to cooperation among users, which avoid to interfere too much with each other. Once the best-response mapping in $\mathcal{OP}_3$ is computed, the solution is combined through a (possibly time-varying) step-size $\alpha^t$ as: 
\begin{equation}
    \bx_q^{t+1}=\bx_q^t+\alpha^t(\widehat{\bx}_q^t-\bx_q^t),
\end{equation} 
for $q=1,\ldots,Q$. The overall procedure, termed as Distributed Successive Concave Approximation (D-SCA), is summarized in Algorithm 1.
\begin{algorithm}[t]
$\textbf{Input}:$ $\tau\geq 0$, $\{\alpha^t\}\geq 0$, $\bx^0_q\in\mathcal{X}_q$, for all $q$. Set $t=0$.\smallskip

\texttt{$\mbox{(S.1)}$}$\,$ If $\bx^t$ satisfies a termination criterion: STOP; \smallskip

\texttt{$\mbox{(S.2)}$}$\,$ For all $q=1,\ldots,Q$, compute $\widehat{\bx}_q^t$ in $\mathcal{OP}_3$;\smallskip

\texttt{$\mbox{(S.3)}$}$\,$ For all $q=1,\ldots,Q$, set:
\begin{equation*}
    \bx_q^{t+1}=\bx_q^t+\alpha^t(\widehat{\bx}_q^t-\bx_q^t);
\end{equation*}

\texttt{$\mbox{(S.3)}$}$\,$ $t\leftarrow t+1$ and go to \texttt{$\mbox{(S.1)}$}.

\caption{D-SCA}
\end{algorithm}

Interestingly, $\mathcal{OP}_3$ (i.e., step \texttt{$\mbox{S.2}$} in Algorithm 1) can be solved distributively by each user $q$ (e.g., by each BS), once the MUI is locally estimated at the $q$-th UE (i.e., the term $\sigma_{q}^2+ \sum_{j=1}^N |H^k_{jq}(\boldsymbol{\phi}^t_{jk})|^2 p^t_{jk} $ in (\ref{eq:Surrogate}) for all $k$ and $q$), and the price vectors $\{\boldsymbol{\pi}^t_{qj}\}_{j\neq q}$ in (\ref{eq:Prices}) are transmitted to user $q$ by all other users $j\neq q$. In particular, $\mathcal{OP}_3$ decouples into two (strongly concave) sub-problems associated with power allocation and RIS optimization, respectively. In the sequel, we illustrate the solution of the two sub-problems.
 \vspace{-0.30cm}
\subsection{Local Power Allocation} \label{sec:power_allocation} 
Solving $\mathcal{OP}_3$ with respect to the power allocation $\bp_q$ leads to the following sub-problem:
\begin{align}\label{eq:SubProblem1}
\mathcal{OP}_4:\quad&\max_{\bp_q}\;\; \sum_{k=1}^K \log_2\left(1+\frac{|H^k_{qq}(\boldsymbol{\phi}^t_{qk})|^2 p_{qk}}{\sigma_{q}^2+ \sum_{j\neq q} |H^k_{jq}(\boldsymbol{\phi}^t_{jk})|^2 p^t_{jk}  }\right)\nonumber\\
&\qquad\qquad+\overline{{\boldsymbol{\pi}}}_q^{t^T} \bp_q  - \frac{\tau}{2}\|\bp_q-\bp_q^t\|^2 \nonumber\\
& \quad \hbox{s.t.} \quad  p_{qk} \geq 0, \,\, \sum_{k=1}^K\;p_{qk} \leq P_q \quad \forall k,q,\nonumber
\end{align}
where $\overline{\boldsymbol{\pi}}^t_q =\{\overline{\pi}^t_{qk}\}_{k=1}^K $ is the part of the pricing vector $\boldsymbol{\pi}^t_q$ in (\ref{eq:Prices}) associated with $\bp_q$, given by:  \vspace{-0.30cm}
\begin{equation}\label{eq:Prices2}
    \overline{\pi}^t_{qk}\,=\,-\sum_{j\neq q}^Q |H^k_{qj}(\boldsymbol{\phi_{qk}^t})|^2 \frac{{\rm snr}^t_{jk}}{(1+{\rm snr}^t_{jk}){\rm MUI}_{jk}^t},
\end{equation}
where ${\rm snr}^t_{jk}$ and ${\rm MUI}_{jk}^t$ are the SINR
and the multi-user interference-plus-noise power experienced by user $j$, generated by the resource allocation profile $\bx^t$:
\begin{align}
    &{\rm snr}^t_{jk}=\frac{|H^k_{jj}(\boldsymbol{\phi}^t_{jk})|^2p_{jk}^t}{{\rm MUI}_{jk}^t}, \\
    &{\rm MUI}_{jk}^t=\sigma_{jk}^2+\sum_{q\neq j} |H^k_{qj}(\boldsymbol{\phi}^t_{qk})|^2 p_{qk}^t. \vspace{-0.25cm}
\end{align} 
The $\mathcal{OP}_4$ can be solved in closed form (up to the multiplier associated with the power budget constraint) and admits the following multi-level water filling solution \cite{scutari2013decomposition}:
\begin{align}\label{eq:Optimal_p}
    &\widehat{\bp}^t_q=\Bigg[ \frac{1}{2}\bp_q^t \circ (1-({\rm \mathbf{snr}_q^t})^{-1}) \nonumber\\
    &- \frac{1}{2\tau}\left( \widetilde{\boldsymbol{\mu}}_q-\sqrt{[\widetilde{\boldsymbol{\mu}}_q-\tau \bp_q^t \circ (1+({\rm \mathbf{snr}_q^t})^{-1})]^2 + 4\tau \mathbf{1} }\right) \Bigg]_+
\end{align}
where $({\rm \mathbf{snr}_q^t})^{-1}=(1/{\rm  snr_{qk}^t})_{k=1}^K$, and $\widetilde{\boldsymbol{\mu}}_q=\overline{\boldsymbol{\pi}}_q^t+\mu_q \mathbf{1}$, where the multiplier $\mu_q$ is chosen to satisfy the complementary condition $0\leq \mu_q \perp \mathbf{1}^T \widehat{\bp}^t_q - P_q \leq 0$. The optimal $\mu_q$ can be efficiently computed (in a finite number of steps) using a bisection method as described in \cite{palomar2005practical}. Note that \eqref{eq:Optimal_p} can be computed efficiently and locally by each user, once the interference generated by the other users (i.e., the  ${\rm MUI}_{jk}^t$ coefficients) and the current interference price $\boldsymbol{\pi}_q^t$ are properly estimated. Of course, the estimation of the prices requires some signaling among users. In practical scenarios, each user interferes only with a subset of "neighbor" users, and thus need to exchange interference prices only with them.
\vspace{-0.20cm}
\subsection{Local RIS Optimization} \label{sec:ris_config}
Solving $\mathcal{OP}_3$ with respect to the $q$-th RIS parameters $\boldsymbol{\phi}_q$ leads to the following sub-problem:
\begin{align}
\mathcal{OP}_5:\quad&\max_{\boldsymbol{\phi}_q}\;\; \Re\{(\boldsymbol{\gamma}_q^t+\underline{\boldsymbol{\pi}}^t_q)^H \boldsymbol{\phi}_q \}  - \frac{\tau}{2}\|\boldsymbol{\phi}_q-\boldsymbol{\phi}_q^t\|^2 \nonumber\\
& \quad \hbox{s.t.} \,\,\,\, \phi_{qk}^m = \frac{F_q^m \omega_k^2}{(\omega_q^m)^2 - \omega_k^2 + \jmath \kappa_q^m \omega_k},\nonumber\\
& \quad \quad \quad |\phi_{qk}^m|\leq 1, \quad \forall k,m,q \nonumber
\end{align}
where $\underline{\boldsymbol{\pi}}^t_q =\{\underline{\pi}^t_{qk}\}_{k=1}^K $ is the part of the pricing vector $\boldsymbol{\pi}^t_q$ in \eqref{eq:Prices} associated with $\boldsymbol{\phi}_q$. The expressions for $\boldsymbol{\gamma}_q^t = \{\boldsymbol{\gamma}^t_{qk}\}_{k=1}^K$ and the pricing vectors $\underline{\boldsymbol{\pi}}^t_q$ are given by: 
\begin{align}
    \boldsymbol{\gamma}^t_{qk} &= \frac{2 p_{qk}^t}{(1 + {\rm snr}^t_{qk}){\rm MUI}^t_{qk}}\mathbf{A}_{qq}\boldsymbol{\phi}_{qk}^t, \label{gamma} \\
    \underline{\pi}^t_{qk} &= -2\sum_{j\neq q}^Q \frac{p_{qk}^t {\rm snr_{jk}^t}}{(1 + {\rm snr}^t_{jk}){\rm MUI}^t_{jk}} \mathbf{A}_{jq} \boldsymbol{\phi}_{qk}^t, \label{PricesPhi} 
\end{align}
where $\mathbf{A}_{jq} \triangleq (\bg_{jq}^* \odot \bh_{qq}^*)(h_{qj}^d + \bg_{jq}^T \odot \bh_{qq}^T)$. To solve $\mathcal{OP}_5$ with respect to the Lorentzian parameters, while enforcing the modulus constraints, we employ the Penalty Dual Decomposition (PDD) method \cite{Shi2020_PDD}, which is suitable for this problem, whose solution can be found in a parallel way. In particular, we may write $\mathcal{OP}_5$'s Augmented Lagrangian (AL) problem by penalizing the equality constraints with the parameter $\rho$ as follows: \vspace{-0.15cm}
\begin{align}
\mathcal{OP}_6:\quad&\min_{\boldsymbol{\phi}_q}\;\; \frac{\tau}{2}\|\boldsymbol{\phi}_q-\boldsymbol{\phi}_q^t\|^2 - \Re\{(\boldsymbol{\gamma}_q^t+\underline{\boldsymbol{\pi}}^t_q)^H \boldsymbol{\phi}_q \} \nonumber\\ 
&\quad\quad +\frac{1}{2\rho} \| \boldsymbol{\phi}_q - \mathbf{d}(\{F_q^m,\omega_q^m, \kappa_q^m \}) + \rho \boldsymbol{\lambda} \|^2 \nonumber\\
& \quad \hbox{s.t.} \,\,\,\, \phi_{qk}^m = \frac{F_q^m \omega_k^2}{(\omega_q^m)^2 - \omega_k^2 + \jmath \kappa_q^m \omega_k},\nonumber\\
& \quad \quad \quad |\phi_{qk}^m|\leq 1, \quad \forall k,m,q, \nonumber
\end{align}
where the vector $\mathbf{d} \in \mathbb{C}^{KM}$, corresponds to the equality constraint in $\mathcal{OP}_5$ related to the Lorentzian phase-response, while $\boldsymbol{\lambda}$ denotes the associated dual variable vector. Then, $\mathcal{OP}_6$ can be solved based on a two-layer procedure, with the inner layer alternatively optimizing $\boldsymbol{\phi}_q$ and the Lorentzian parameters, while the outer layer updating $\rho$ and $\boldsymbol{\lambda}$.

In the inner layer, for a given $\mathbf{d}$, the optimal $\boldsymbol{\phi}_q$ can be computed in closed form and is given by:
\begin{equation}
    \widehat{\boldsymbol{\phi}}_q = \mathcal{P}\left( \frac{1}{\tau + \frac{1}{\rho}} \left( \tau \boldsymbol{\phi}_q^t+ (\boldsymbol{\gamma}_q^t+\underline{\boldsymbol{\pi}}^t_q) + \frac{1}{\rho}(\mathbf{d} - \rho \boldsymbol{\lambda}) \right) \right),
\end{equation}
where $\mathcal{P}(\cdot)$ is a component-wise operator applied to complex entries, such that: \vspace{-0.40cm}
\begin{equation}
    \mathcal{P}(y)=\begin{cases} 
    y, &\mbox{if } |y|\leq 1, \\
    y/|y|, & \mbox{if } |y|> 1. \end{cases}
\end{equation}
Then, for the obtained $\widehat{\boldsymbol{\phi}}_q$, $\mathcal{OP}_6$ reduces to the following non-linear least squares sub-problem:
\begin{equation}
    \{\hat{F}_q^m,\hat{\omega}_q^m, \hat{\kappa}_q^m \} = \mathop{\arg\min}\limits_{\{F_q^m, \omega_q^m, \kappa_q^m\}_{m=1}^M} \| \mathbf{s} - \mathbf{d}(\{F_q^m,\omega_q^m, \kappa_q^m \}) \|^2,
\end{equation}
where $\mathbf{s} \triangleq \widehat{\boldsymbol{\phi}}_q + \rho \boldsymbol{\lambda}$, which can be solved by employing the Levenberg-Marquardt algorithm. Then, in the outer layer, $\boldsymbol{\lambda}$ and $\rho$ are updated by $\boldsymbol{\lambda} \leftarrow \boldsymbol{\lambda} + \rho^{-1}(\widehat{\boldsymbol{\phi}}_q - \hat{\mathbf{s}})$ and $\rho \leftarrow c \rho$, where $c<1$ is a constant scaling factor. The proposed overall PDD-based algorithm is omitted here due to space limitations, but the reader is referred to \cite[Alg. 1]{katsanos2022wideband} for details.
\vspace{-0.05cm}
\section{Numerical Results} \label{sec:num_results} \vspace{-0.10cm}
\subsection{Experimental Setup} \label{sec:exp_setup} \vspace{-0.15cm}
In this section, we evaluate the performance of the proposed distributed design of power allocation and RIS configuration for each user of the considered system. We have considered the Rayleigh fading channels $\{\bar{h}_{qj}^d[i]\}_{i=0}^{L-1}$, $\{\bar{\bg}_{qj}[i]\}_{i=0}^{L-1}$, and $\{\bar{\bh}_{qq}[i]\}_{i=0}^{L-1}$ in the time domain, where $L$ denotes the number of delayed taps in the channel impulse response for each link. Then, the frequency-domain channel for the link $h_{qj}^d$ at each $k$-th SC was obtained as $h_{qj}^d[k] = \sum_{i=0}^{L-1} \bar{h}_{qj}^d[i] e^{- \jmath 2\pi (k-1)i/K}$; we similarly derived all considered channels. We also assumed that each channel consists of independent entries with zero mean and unit variance, which were multiplied by distance dependent pathloss that was modeled as $\operatorname{PL}_{mn} = \operatorname{PL}_0 (d_{mn}/d_0)^{-\alpha_{mn}}$, with $\operatorname{PL}_0 = -30$dB denoting the pathloss at the reference distance $d_0 = 1\,m$, and $\alpha_{mn}$ being the pathloss exponent for the channel between nodes $m$ and $n$ with distance $d_{mn}$. The pathloss exponents for the RIS involved channels were set equal to $2$ and $4$ for the rest of them. In our simulations, we first assumed the number of users being equal to $Q = 2$. In this scenario, BS1 was placed at the origin of the $xy$-plane, and BS2 was located in the point with coordinates $(2d,0)$, where $d = 20\,m$. For the remaining nodes, that is, RIS1, UE1, RIS2, and UE2, we considered the fixed positions with coordinates: $(-d/4,d/8)$, $(d/2,3d/2)$, $(9d/4,d/8)$, and $(3d/2,3d/2)$, respectively. For the case of $Q = 3$ users, we considered the previous locations and the placements of BS3, RIS3 and UE3 were accordingly at the points $(0,4d)$, $(-d/4,9d/4)$, and $(d/2,5d/2)$. In addition, we considered $M=50$ RIS unit elements, $K = 16$ SCs, $L = 4$ taps, and $\sigma_{q}^2= -80$dBm $\forall$$q$. The Lorentzian parameters were constrained as follows: $F_n \in (0,1]$, $\omega_n > 0 $ and $\lvert \kappa_n \rvert \leq 100$, while $\lvert \omega \rvert \leq \pi$, while the threshold for the termination criterion of the proposed algorithm was set as $\epsilon = 10^{-3}$. The achievable sum-rate performance results were averaged over $100$ independent Monte Carlo realizations.

\subsection{Sum-Rate Performance} \label{sec:results_sum_rate} 
\begin{figure}[!t]
	\centering
		\includegraphics[scale=0.60]{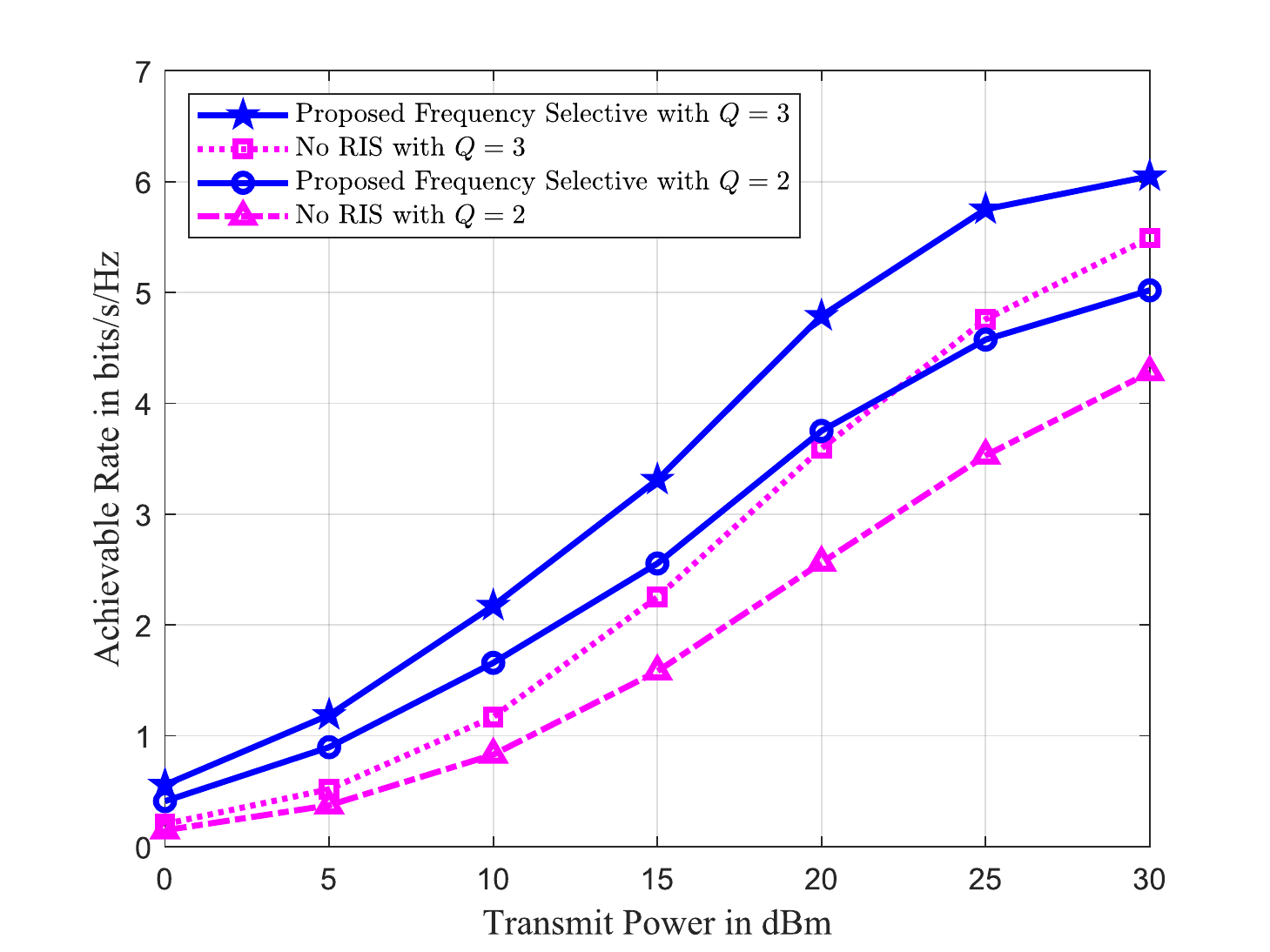}
	\caption{Achievable sum-rate performance in bps/Hz as a function of the transmit power $P_q=P$ in dBm for each BS, considering the proposed distributed design for frequency-selective RISs with $M=50$ unit elements, OFDM transmission with $K = 16$ SCs, and different number of users, $Q$.}
\label{fig:Rate_vs_TX_Power}
\end{figure}

\vspace{-0.10cm}
We investigate in Fig$.$~\ref{fig:Rate_vs_TX_Power} the achievable sum rate in bps/Hz of the proposed distributed algorithm versus the total transmit power $P_q = P$ at each $q$-th BS for different numbers $Q$ of users. For comparison purposes, we have implemented the proposed D-SCA Algorithm for the case where no RISs are present. It can be observed that, for both evaluated schemes, the total sum rates increase with $P$, whereas in the high SNR regime (i.e., above $20$ dBm), a saturation trend is present. This behavior is expected due to the fact that interference from neighboring users is inevitable. In addition, as illustrated, inserting one RIS for each user, even with the relatively small number $M=50$ of unit elements, leads to higher sum rate, confirming the benefits of the RIS technology.
 \vspace{-0.20cm}
\section{Conclusions} \label{sec:conclusions} \vspace{-0.20cm}
In this paper, we studied RIS-empowered cellular OFDM wireless systems comprised  many BS-RIS-UE triplets, where each RIS unit element acts as a resonant circuit modeled by a Lorentzian-like frequency response. To characterize system's performance, we adopted the achievable sum-rate metric and formulated a maximization problem for the BS power allocation and the multi-RIS configuration, which was solved based on a distributed optimization approach. Through numerical evaluations, it was shown that utilizing one RIS for each BS-UE pair leads to significant gains in contrast to the case without RISs, even when a small-sized RIS is deployed.

\vspace{-0.20cm}
\bibliographystyle{IEEEtran}
\bibliography{IEEEabrv, refs}

\begin{thebibliography}{10}
\providecommand{\url}[1]{#1}
\csname url@samestyle\endcsname
\providecommand{\newblock}{\relax}
\providecommand{\bibinfo}[2]{#2}
\providecommand{\BIBentrySTDinterwordspacing}{\spaceskip=0pt\relax}
\providecommand{\BIBentryALTinterwordstretchfactor}{4}
\providecommand{\BIBentryALTinterwordspacing}{\spaceskip=\fontdimen2\font plus
\BIBentryALTinterwordstretchfactor\fontdimen3\font minus
  \fontdimen4\font\relax}
\providecommand{\BIBforeignlanguage}[2]{{%
\expandafter\ifx\csname l@#1\endcsname\relax
\typeout{** WARNING: IEEEtran.bst: No hyphenation pattern has been}%
\typeout{** loaded for the language `#1'. Using the pattern for}%
\typeout{** the default language instead.}%
\else
\language=\csname l@#1\endcsname
\fi
#2}}
\providecommand{\BIBdecl}{\relax}
\BIBdecl

\bibitem{huang2019reconfigurable}
C.~Huang \emph{et~al.}, ``Reconfigurable intelligent surfaces for energy
  efficiency in wireless communication,'' \emph{IEEE Trans. Wireless Commun.},
  vol.~18, no.~8, pp. 4157--4170, 2019.

\bibitem{di2019smart}
M.~Di~Renzo~et al., ``Smart radio environments empowered by reconfigurable {AI}
  meta-surfaces: An idea whose time has come,'' \emph{EURASIP J. Wireless
  Commun. Netw.}, vol.~1, pp. 1--20, 2019.

\bibitem{RISE6G_COMMAG}
E.~Calvanese~Strinati \emph{et~al.}, ``Reconfigurable, intelligent, and
  sustainable wireless environments for {6G} smart connectivity,'' \emph{IEEE
  Commun. Mag.}, vol.~59, no.~10, pp. 99--105, Oct. 2021.

\bibitem{RIS_Overview}
M.~Jian \emph{et~al.}, ``Reconfigurable intelligent surfaces for wireless
  communications: {O}verview of hardware designs, channel models, and
  estimation techniques,'' 2022, [Online] https://arxiv.org/pdf/2203.03176.pdf.

\bibitem{Samarakoon_2020}
G.~C. Alexandropoulos \emph{et~al.}, ``Phase configuration learning in wireless
  networks with multiple reconfigurable intelligent surfaces,'' in \emph{Proc.
  IEEE GLOBECOM}, Taipei, Taiwan, Dec. 2020.

\bibitem{Huang2021MultiHop}
C.~Huang \emph{et~al.}, ``Multi-hop {RIS}-empowered terahertz communications:
  {A DRL}-based hybrid beamforming design,'' \emph{IEEE J. Sel. Areas Commun.},
  vol.~39, no.~6, pp. 1663--1677, Jun. 2021.

\bibitem{Stylianopoulos2022DeepCB}
K.~Stylianopoulos \emph{et~al.}, ``Deep contextual bandits for orchestrating
  multi-user {MISO} systems with multiple {RISs},'' in \emph{IEEE ICC}, Seoul,
  South Korea, May 2022, [Online] https://arxiv.org/pdf/2202.08194.pdf.

\bibitem{yang2020_resource}
Z.~Yang \emph{et~al.}, ``Resource allocation for wireless communications with
  distributed reconfigurable intelligent surfaces,'' in \emph{Proc. IEEE
  GLOBECOM}, dec. 2020.

\bibitem{Marco21_multiRISMISO}
M.~He \emph{et~al.}, ``Cooperative multi-{RIS} communications for wideband
  mmwave {MISO-OFDM} systems,'' \emph{IEEE Wireless Commun. Lett.}, vol.~10,
  no.~11, pp. 2360--2364, Nov. 2021.

\bibitem{an2021}
J.~An \emph{et~al.}, ``Reconfigurable intelligent surface-enhanced {OFDM}
  communications via delay adjustable metasurface,'' Oct. 2021, [Online]
  https://arxiv.org/abs/2110.09291.

\bibitem{katsanos2022wideband}
K.~D. Katsanos \emph{et~al.}, ``Wideband multi-user {MIMO} communications with
  frequency selective {RIS}s: Element response modeling and sum-rate
  maximization,'' in \emph{Proc. IEEE ICC}, Seoul, South Korea, May 2022.

\bibitem{Shlezinger_2021_DMAs}
N.~Shlezinger \emph{et~al.}, ``Dynamic metasurface antennas for 6{G} extreme
  massive {MIMO} communications,'' \emph{IEEE Wirel. Commun.}, vol.~28, no.~2,
  pp. 106--113, Apr. 2021.

\bibitem{DSmith-2017PRA}
D.~R. Smith \emph{et~al.}, ``Analysis of a waveguide-fed metasurface antenna,''
  \emph{Phys. Rev. Appl.}, vol.~8, no.~5, pp. 1--16, Nov. 2017.

\bibitem{Zhang_2020_MIMO}
S.~Zhang and R.~Zhang, ``Capacity characterization for intelligent reflecting
  surface aided {MIMO} communication,'' \emph{IEEE J. Sel. Areas Commun.},
  vol.~38, no.~8, pp. 1823--1838, Aug. 2020.

\bibitem{scutari2013decomposition}
G.~Scutari \emph{et~al.}, ``Decomposition by partial linearization: Parallel
  optimization of multi-agent systems,'' \emph{IEEE Trans. Signal Process.},
  vol.~62, no.~3, pp. 641--656, Feb. 2014.

\bibitem{facchinei2015parallel}
F.~Facchinei \emph{et~al.}, ``Parallel selective algorithms for nonconvex big
  data optimization,'' \emph{IEEE Trans. Signal Process.}, vol.~63, no.~7, pp.
  1874--1889, Apr. 2015.

\bibitem{palomar2005practical}
D.~P. Palomar and J.~R. Fonollosa, ``Practical algorithms for a family of
  waterfilling solutions,'' \emph{IEEE Trans. Signal Process.}, vol.~53, no.~2,
  pp. 686--695, Feb. 2005.

\bibitem{Shi2020_PDD}
Q.~Shi and M.~Hong, ``Penalty dual decomposition method for nonsmooth nonconvex
  optimization \textemdash{{Part I}}: Algorithms and convergence analysis,''
  \emph{IEEE Trans. Signal Process.}, vol.~68, pp. 4108--4122, Jun. 2020.

\end{thebibliography}

\end{document}